\documentclass[preprint,12pt]{elsarticle}

\usepackage{amssymb}
\usepackage{lineno}
\begin{document}

\begin{frontmatter}

\title{Slow Control Systems of the Reactor Experiment for Neutrino Oscillation}
%\linenumbers
\author[dsu]{J.~H. Choi}
\author[syu]{H.~I. Jang}
\author[snu]{W.Q. Choi}
\author[skku]{Y. Choi}
\author[gist]{J.~S. Jang}
\author[ibs]{E.~J. Jeon}
\author[cnu]{K.~K. Joo}
\author[cnu]{B.~R. Kim}
\author[sju]{H.~S. Kim}
\author[cnu]{J.~Y. Kim} 
\author[snu]{S.~B. Kim}
\author[pnu]{S.~Y. Kim}
\author[knu]{W. Kim}
\author[ibs]{Y.~D. Kim}
\author[cau]{Y.~J. Ko}
\author[pnu]{J.~K. Lee}
\author[cnu]{I.~T. Lim}
\author[dsu]{M.~Y. Pac\corref{cor1}}
\ead{pac@dsu.kr}
\cortext[cor1]{Corresponding author}
\author[gnu]{I.~G. Park}
\author[snu]{J.~S. Park}
\author[cnu]{R.~G. Park}
\author[snu]{H.~K. Seo}
\author[snu]{S.~H. Seo}
\author[cnu]{C.~D. Shin}
\author[cau]{K. Siyeon}
\author[cnu]{I.~S. Yeo}
\author[skku]{I. Yu}

\address[cnu]{Institute for Universe \& Elementary Particles, Chonnam National University, Gwangju, 61186, Korea}
\address[cau]{Department of Physics, Chung-Ang University, Seoul, 06974, Korea}
\address[dsu]{Basic Science Research Institute, Dongshin University, Naju, 58245, Korea}
\address[gist]{GIST College, Gwangju Institute of Science and Technololgy, Gwangju 61005, Korea} 
\address[gnu]{Department of Physics, Gyeongsang National University, Jinju, 52828, Korea}
\address[knu]{Department of Physics, Kyungpook National University, Daegu, 41566, Korea}
\address[sju]{Department of Physics and Astronomy, Sejong University, Seoul, 05006, Korea}
\address[snu]{Department of Physics \& Astronomy, Seoul National University, Seoul, 08826, Korea}
\address[syu]{Department of Fire Safety, Seoyeong University, Gwangju, 61268, Korea}
\address[skku]{Department of Physics, Sungkyunkwan University, Suwon, 16419, Korea}

\begin{abstract}
The RENO experiment has been in operation since August 2011 to measure reactor antineutrino disappearance using identical near and far detectors. For accurate measurements of neutrino mixing parameters and efficient data taking, it is crucial to monitor and control the detector in real time. Environmental conditions also need to be monitored for stable operation of detectors as well as for safety reasons. In this article, we report the design, hardware, operation, and performance of the slow control system.
\end{abstract}


\end{frontmatter}

%\linenumbers
%% main text
\section{Introduction}
The Reactor Experiment for Neutrino Oscillation (RENO) has successfully measured the value of the smallest neutrino mixing angle, $\theta_{13}$, and has undertaken the measurement of the squared mass difference $|\Delta m^{2}_{ee}|$~\cite{jkahn1, sbk}.  The inverse beta decay data collected at RENO uses electron antineutrinos produced by six equally spaced reactors of the Hanbit nuclear power plant in Yeonggwang, Korea. There are two identical detectors located at near and far sites at 294 m and 1,383 m, respectively, from the center of the reactor array. The power plant consists of six equally spaced reactor cores placed linearly and provides a total thermal power of 16.8 $\rm{GW}_{th}$~\cite{jkahn2}. 

The RENO detector consists of a main inner detector (ID) and an outer veto detector (OD). The ID is contained in a cylindrical stainless steel vessel that houses two concentric cylindrical acrylic vessels. The innermost vessel is filled with $18.3 ~{\rm m^{3}} (\sim 16 ~{\rm tons})$ of $\sim0.1$\% Gd-loaded liquid scintillator (LS) as a neutrino target. This innermost vessel is surrounded by a $\gamma$-catcher region with a 60-cm-thick layer of unloaded LS inside an outer acrylic vessel. Outside the $\gamma$-catcher is a 70-cm-thick buffer region filled with 65 tons of mineral oil~\cite{isyeo,kspark,jspark}. On the inner wall of the stainless steel vessel, 354 10-inch Hamamatsu R7081 photomultiplier tubes (PMTs) are mounted.   Enclosing the ID, an 1.5-m-thick region of the outer detector is filled with highly purified water and 67 R7081 PMTs are mounted on the wall of the OD vessel for vetoing cosmic ray muons. 

For accurate measurements of  $\theta_{13}$ and  $|\Delta m^{2}_{ee}|$, the two identical detectors must be operated in stable and similar conditions, and these detectors must be monitored in real time. The RENO slow control systems (SCS) are responsible for the verification of reliable operation and prompt monitoring of experimental conditions.

This article presents the design, installation, operation, and performance of the SCS. Section 2 provides an overview of SCS, describing each SCS component and their performance. We summarize the SCS in Section 3.

\section{Slow Control Systems for RENO}
\subsection{Overview}
The RENO SCS are intended for reliable control and quick monitoring of the detector operational status and environmental conditions. If SCS issues an alarm with a possible deviation from the predefined experimental parameter range, immediate maintenance is required in an automatic procedure or an operator-driven action.

The SCS design and construction enables simultaneous control and monitoring of the two identical near and far detectors in a single location. The central control room, which is located at the entrance of the access tunnel for the far detector, controls and monitors the operation of both detectors through the use of the virtual network computing (VNC) viewer {\sc Teamviewer}. {\sc Teamviewer} is a graphical desktop sharing system based on the remote frame buffer (RFB) protocol for controlling another computer remotely. Figure 1 shows the layout of the RENO experiment site and the SCS adopted for the experiment.
\begin{figure}[htbp]
\centering
\includegraphics[width=130mm]{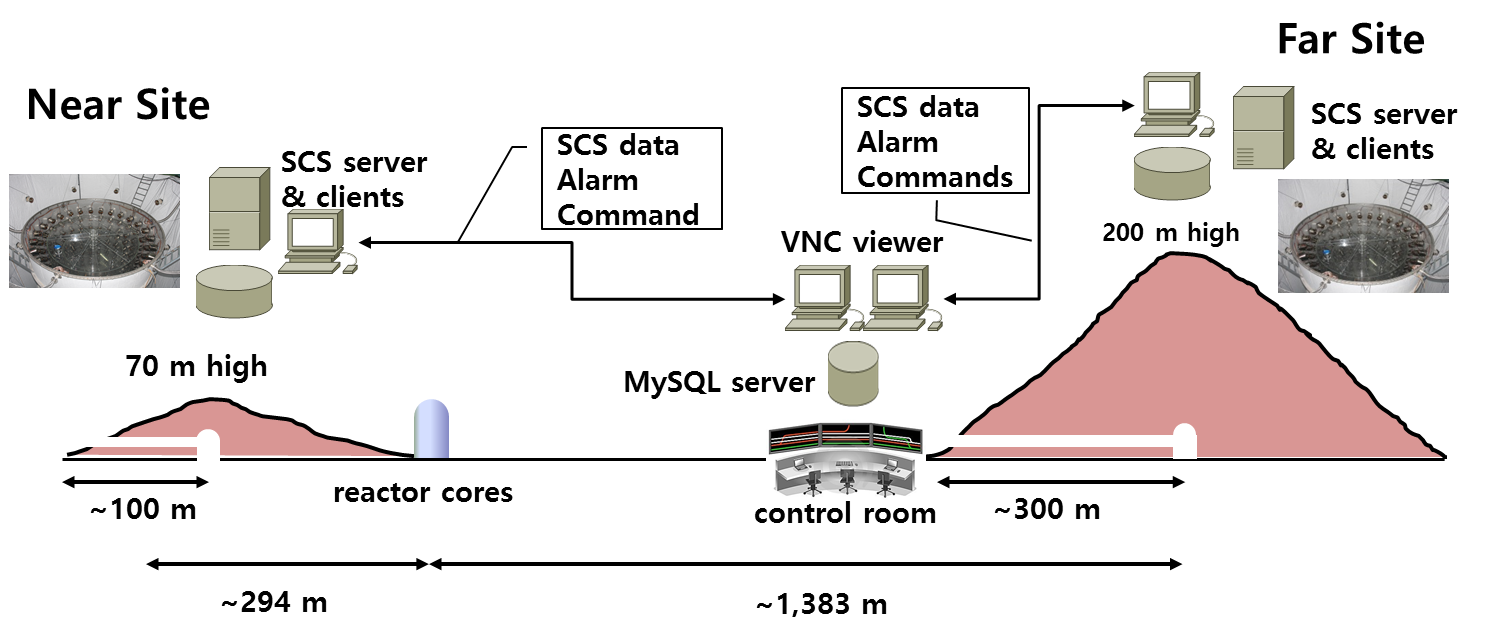}
\caption{The layout of the RENO experiment site and the SCS. The SCS server and clients are installed in an on-site control room inside the access tunnel. The data from the SCS sensors are monitored by the server and delivered to the clients. The clients register and display the information from the server. The storage for the SCS data is installed in the central control room. The VNC viewer enables remote control and monitoring of detector operations.}
\end{figure}

The primary function of SCS is monitoring and control of the high voltage applied to PMTs and the water level in OD. In addition, environmental conditions, such as air quality, temperature, and humidity, in the experimental halls are monitored for safety and protection of electronics.

The acquired sensor data forwarded to SCS client applications for monitoring and storage. Based on the sensor data, the SCS clients display detector status and environmental parameters, generates alarms, and issue control signals.
This system can be accessed through the network to display or record both online and historical data. The SCS client applications allow users to manage and access the status of the experiment through a flexible graphical-user-interface based tool. The schematic drawing of RENO SCS is given in Fig. 2.
\begin{figure}[htbp]
\centering
\includegraphics[width=130mm]{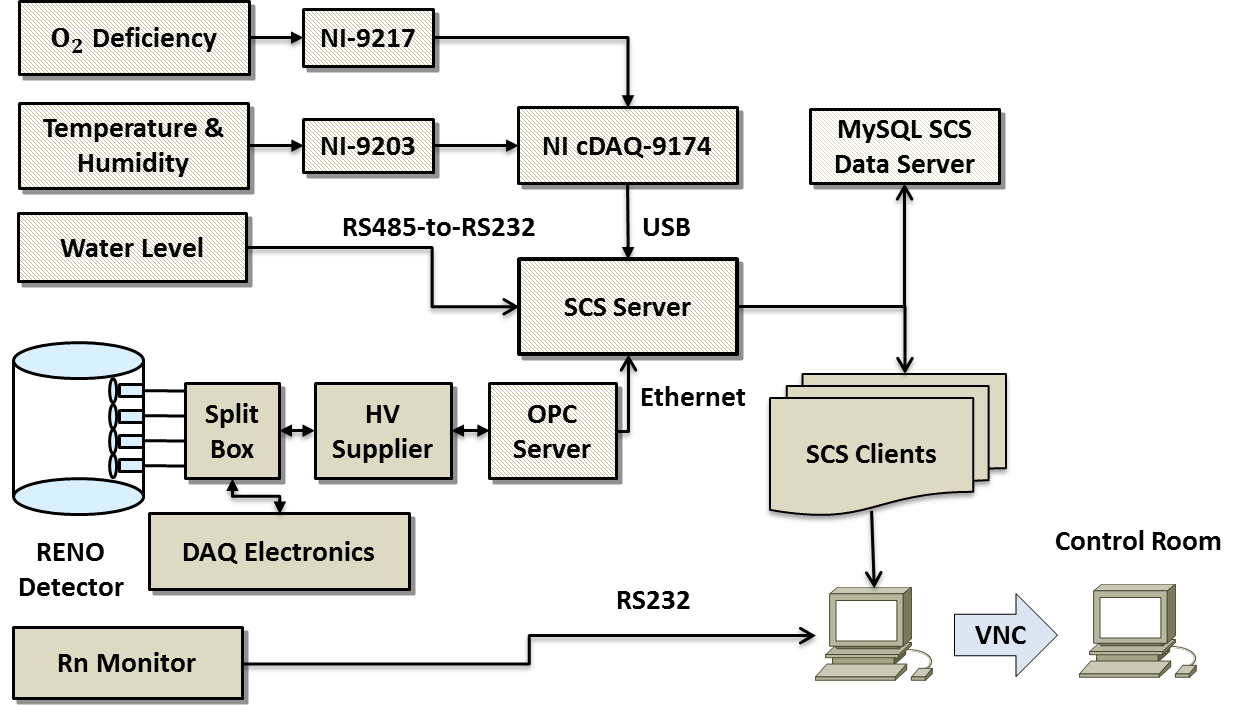}
\caption{Conceptual diagram of the RENO SCS systems. A VNC viewer mirrors the monitoring PC screen to a PC in the central control room.}
\end{figure}

\subsection{Components of the Slow Control Systems and their performance}
The RENO SCS are developed based on a client-server architecture. A server is located at each detector site, and the servers are responsible for the display and storage of all data from the high voltage (HV) crates and sensors. The core of the SCS hardware is implemented on a commercial personal computer (PC) under the {\sc Windows} operating system at each detector site. The majority of the sensors used in RENO use universal serial bus (USB) associated with the RS232 protocol. {\sc Mysql} has been selected as the SCS data storage server, which is installed in the central control room, because {\sc Mysql} is a powerful open-source relational database system that is highly scalable both in the sheer quantity of manageable data and in the number of concurrent users it can accommodate. 

At each detector site, the SCS server running on the local PC records the data and provides the data to the client applications of the online monitoring system. The client applications need to be able to register and display the measurements from the detector. Two types of client applications, HV status panel and environmental condition including OD water level panel, are implemented using {\sc Labview} from National Instruments. The advantage of using {\sc Labview} for this medium-size experiment lies mostly on its intuitive graphical interface and lower dependence on device libraries~\cite{LabVIEW}. These two client applications can query both current and recorded data and are able to send commands to each detector site. The data provided by the SCS server are plotted in real-time on the programmable graphical panels from {\sc Labview}, as shown in Fig. 3. The status and performance of the detector operations are reported based on the online monitoring system during data collection. The RENO SCS transmits the data obtained from each sensor to client applications every 5 ${\rm s}$, except for the HV and radon data. The HV object linking and embedding (OLE) for process control (OPC) server provides the current applying voltages every second, and the radon concentration is recorded every hour.

\begin{figure}[htbp]
\centering
\includegraphics[width=130mm]{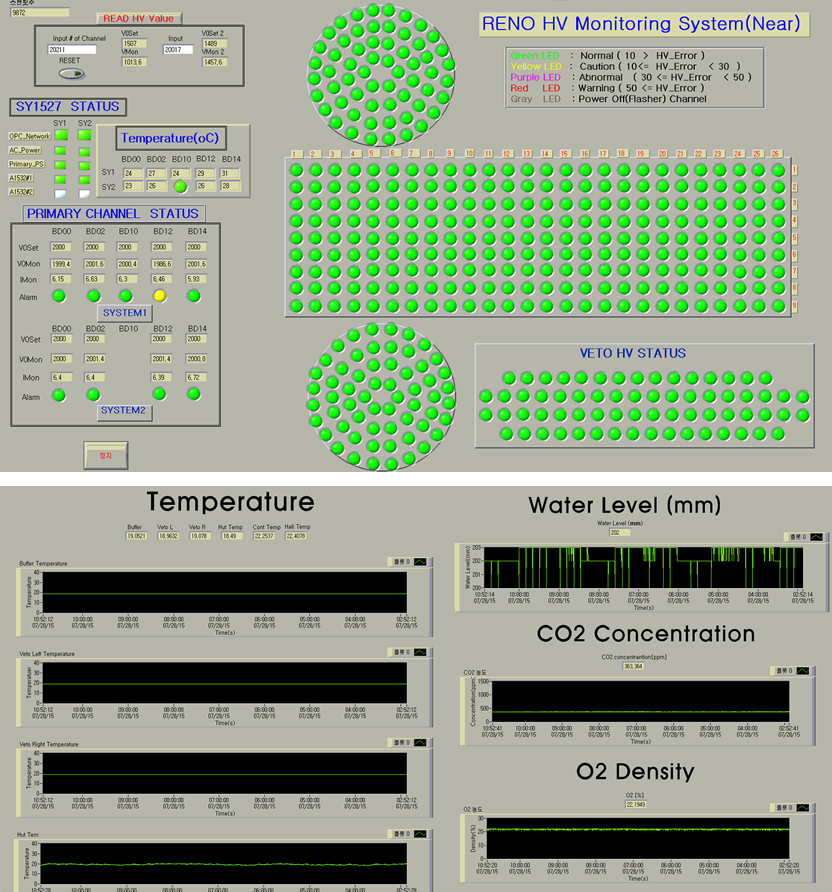}
\caption{Monitoring screens are provided by SCS clients. The upper part of the figure shows the current status of the HV system. The status of the HV applied to each PMT is represented by a color code on a circle. The lower part of the figure shows the current SCS values from the SCS server using numbers in the small boxes, and the history of each SCS channel are shown on the plots.}
\end{figure}

The following parameters are controlled and monitored for ensuring detector integrity, stability, and safety: HV supply, OD water level, liquid temperatures, oxygen deficiency, humidity and temperature of the operating system, and radon concentration. 

\subsubsection{High Voltage Supply} 
The 421 PMTs in a detector are divided into nine sets in the HV supplying system. Each set of 48 PMT channels is connected to a decoupler box, which decouples the HV and pulse signal from the anode using capacitors. Moreover, each set of PMT channels receives the HV from a single module of CAEN A1932. The CAEN SY1527LC main frame provides HV for all 421 PMTs and periodically reports the operating values of HV. The SCS is able to set and read the voltages, monitor tripping conditions, time-stamp the data, and issue alarms if the HV channel deviates from its configured preset. The connection between the HV system and SCS server is based on an OPC server supplied by CAEN ~\cite{caen}. 

The OPC server offers ``plug and play'' connectivity between disparate hardware devices. In general, the introduction of the OPC interface has reduced the number of driver developments, which hardware manufacturers implement for their components, to one driver: the OPC server. On the other hand, OPC client applications (from any vendor) can communicate with the OPC server to exchange data in a standard way. Each device property is accessed via an OPC item. The OPC server creates OPC items to which an OPC client connects. In the client, the OPC items can be organized in groups with a hierarchical structure. In RENO, the OPC server simplifies the client software for controlling and monitoring HV and enables uncomplicated improvements or modifications.

An efficient way to monitor the HV status of the 421 PMTs in each detector is to display a planar figure of the PMT mounted ID vessel, as shown in Fig. 3. The individual HV status is presented by a color code on a circle. The color is determined comparing the current and preset HV values. Green indicates that the difference between the current and preset values is less than 10 V. Yellow corresponds to HV differences within the range from 10 to 30 V. If the difference exceeds 30 V, the color turns to red. Powered-off PMTs are shown in gray. A mouse click on a circle opens the PMT HV history viewer. Fig. 4 shows the voltage history of a failed PMT.
\begin{figure}[htbp]
\centering
\includegraphics[width=140mm]{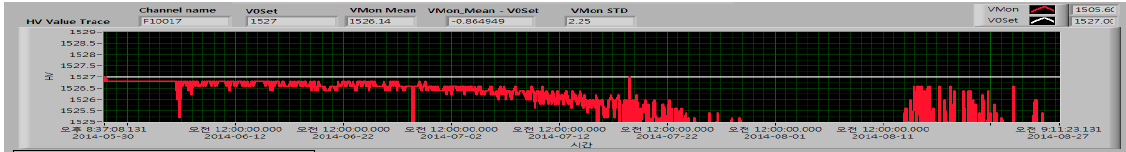}
\caption{History of HV applied to a PMT provided by time stamp.}
\end{figure}

\subsubsection {Water Level of OD} 
The RENO OD is filled with water to operate as a water Cherenkov detector and is used for identifying cosmic muons and for shielding against background radiation coming from the surrounding rock. An 1.5-m-thick layer surrounds the ID. A water purification system was designed to produce 10 tons of pure 18 $\rm{M\Omega}$ water per day. In order to maintain uniform water quality, the purified water is circulated by a water circulation system that also maintains the OD water level.

The water level sensor in RENO is a SICK UM-30 ultrasonic rangefinder, which uses sound pulses to measure distance. The water level sensor is located at the top of the OD vessel and can measure the distance between the water level and top of the OD in millimeters. The distance obtained from the sensor is read through the RS485-to-RS232 bridge interface. The water circulation system is invoked by the water level sensor to maintain a preset water level.

In the last three years of operation, the water level of OD has been maintained within $\sim 2$~mm from the preset water level. The water circulation system, including the client application, is designed to allow for a 2-mm variation of the water level. Figure 5 shows plots of the water levels from the far and near detectors.

\begin{figure}[htbp]
\centering
\includegraphics[width=130mm]{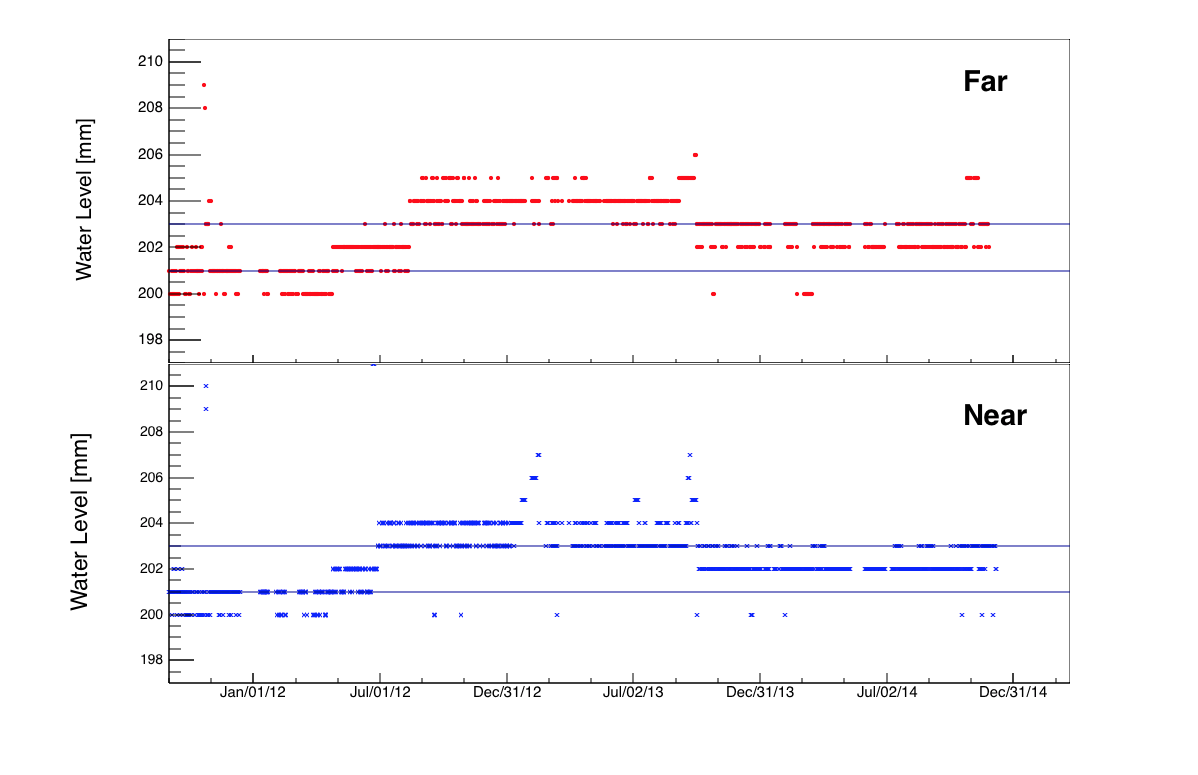}
\caption{Water level from the veto top at the far and near detector. Except for system maintenance periods, the OD water levels have been well controlled within the designed regions shown as between two dotted lines.}
\end{figure}

\subsubsection {Temperature of Liquids} 
The temperature variation may affect the properties of LS and PMT performance. Therefore, monitoring temperature of the detectors is important to understand the detector performance as well as the possible systematic uncertainties stemming from temperature difference between two detectors.
Each detector has two PT-100 thermo-couples on the veto wall and a PT-100 thermo-couple on the buffer wall. The thermo-couples were tested for  uniformity and stability before installation. The thermo-couples are connected to an 8-channel NI-9203 electric current input module, which is digitized by cDAQ NI-9174 up to a 200-kHz sampling rate\cite{LabVIEW}. 

Figure 6 depicts the detector temperatures from the far and near detectors and the relative differences. The difference of the temperatures is maintained within ~3K. On average, the liquid temperatures in the near detector are slightly lower than those of the far detector. It could be due to the access tunnel to near detector being much shorter than that of far detector. 
\begin{figure}[htbp]
\centering
\includegraphics[width=130mm]{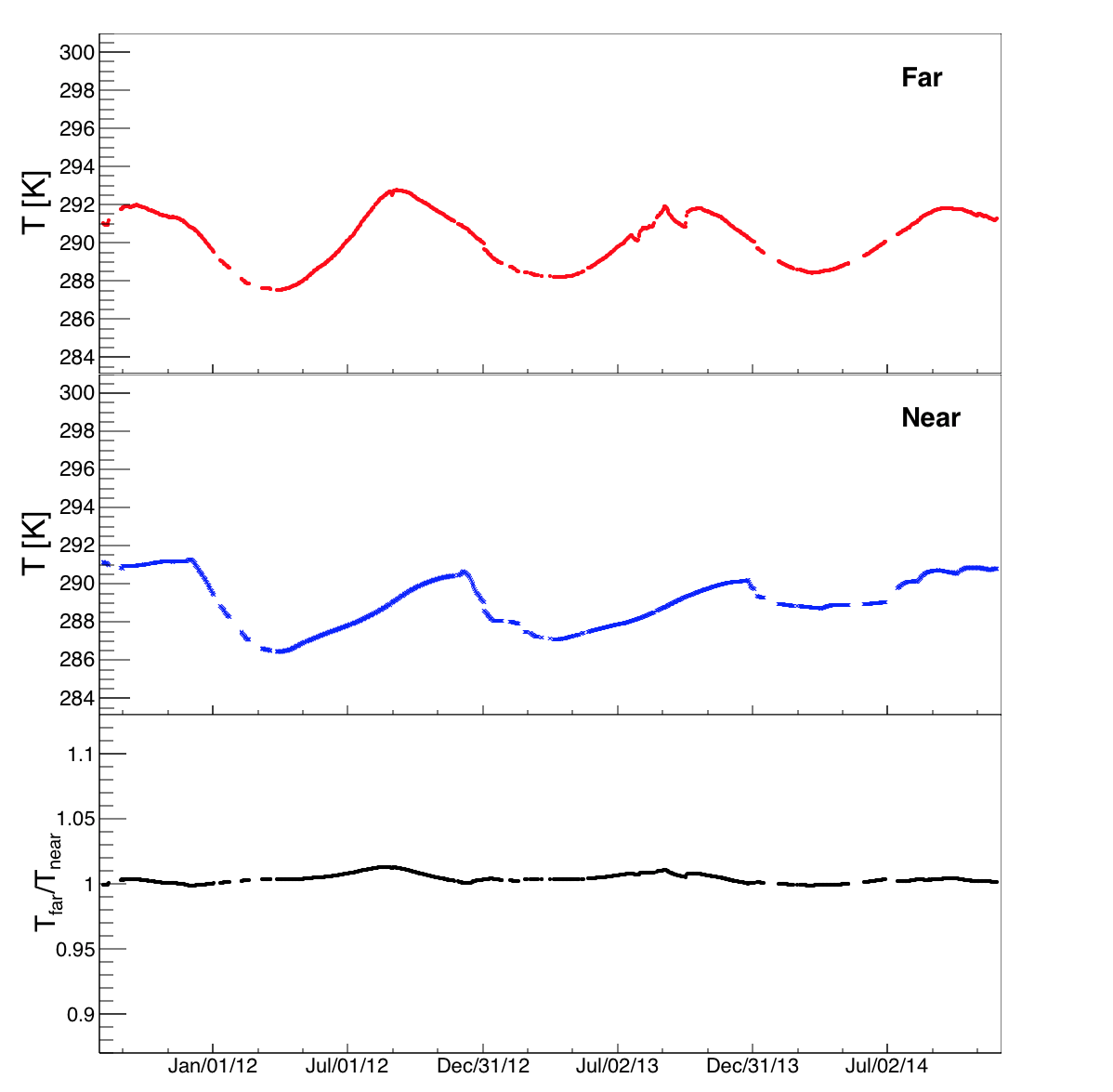}
\caption{Detector temperatures and relative differences obtained from August 1, 2011 to November 19, 2014. Blanks in histogram arise from SCS maintenance or failure. The first and second figures from the top are the detector temperatures from the far and near detectors in Kelvin, respectively. The ratio between the far and near detector, ${\rm T_{far}/T_{near}}$, is plotted in the third figure. During the operation, the relative differences have been maintained below 1\%. }
\end{figure}

\subsubsection {Oxygen Deficiency} 
Concentrations of CO$_2$ and O$_2$ in the experimental hall are monitored using an infrared (IR) absorption technique. The Sense Cube KCD-HP500 for CO$_2$ and the Honeywell Sensepoint XCD for O$_2$ are installed in the on-site control rooms of the near and far experimental halls. The sensors periodically read the value of the CO$_2$ and O$_2$ concentrations and send the results to a database. The sensors are connected to a 4-channel NI-9217 voltage input module, which is digitized by NI-9174 up to a 400-Hz sampling rate.

Figure 7 shows the CO$_2$ and O$_2$ concentrations in the experimental halls from September 2011 to February 2015. According to the safety regulations, the CO$_2$ concentration should be lower than 1,000 ppm, and the O$_2$ concentration should be in the range of 18.0--23.5\%. A ventilation system has been installed inside the tunnel to prevent oxygen deficiency.

\begin{figure}[htbp]
\centering
\includegraphics[width=130mm]{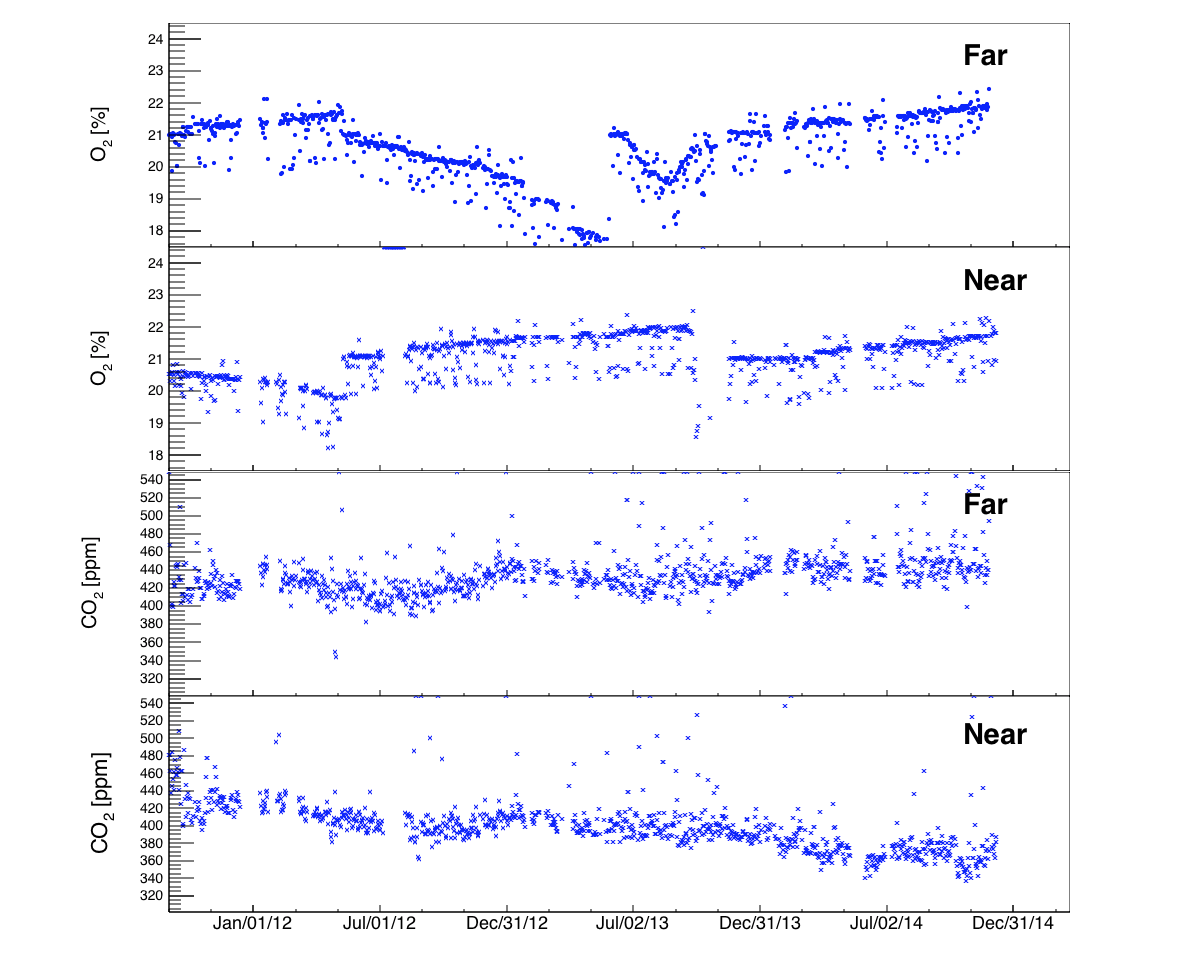}
\caption{Variation of O$_2$ and CO$_2$ concentrations during operation periods. The O$_2$ concentration at the far detector site in the middle of 2013 was relatively low due to failure of the O$_2$ sensor. The CO$_2$ concentrations at far and near detector have been kept below 540 ppm during operation. Tunnel access for maintenance causes relatively high CO$_2$ concentration.}
\end{figure}

\subsubsection {Humidity and Temperature of Electronics System} 
The humidity and temperature of an electronics hut should be maintained at stable levels for the data acquisition (DAQ) and PMT HV systems. The HTM420R integrated device from Dayeon Electronics is installed to measure the temperature and humidity in the electronics hut, the experimental hall, and the on-site control room. The integrated device is connected to a NI-9217 voltage input module.

Figure 8 shows the variation of the temperature and the humidity in the electronics hut during an operation period. To minimize accidental breakdown or unexpected errors of the DAQ electronics and HV system, the humidity and temperature have been cautiously monitored. Using an air conditioner and dehumidification system, the humidity has been maintained in the 10--60\% range, and the temperature has been maintained below $\sim$$25^{\circ}{\rm C}$.

\begin{figure}[htbp]
\centering
\includegraphics[width=130mm]{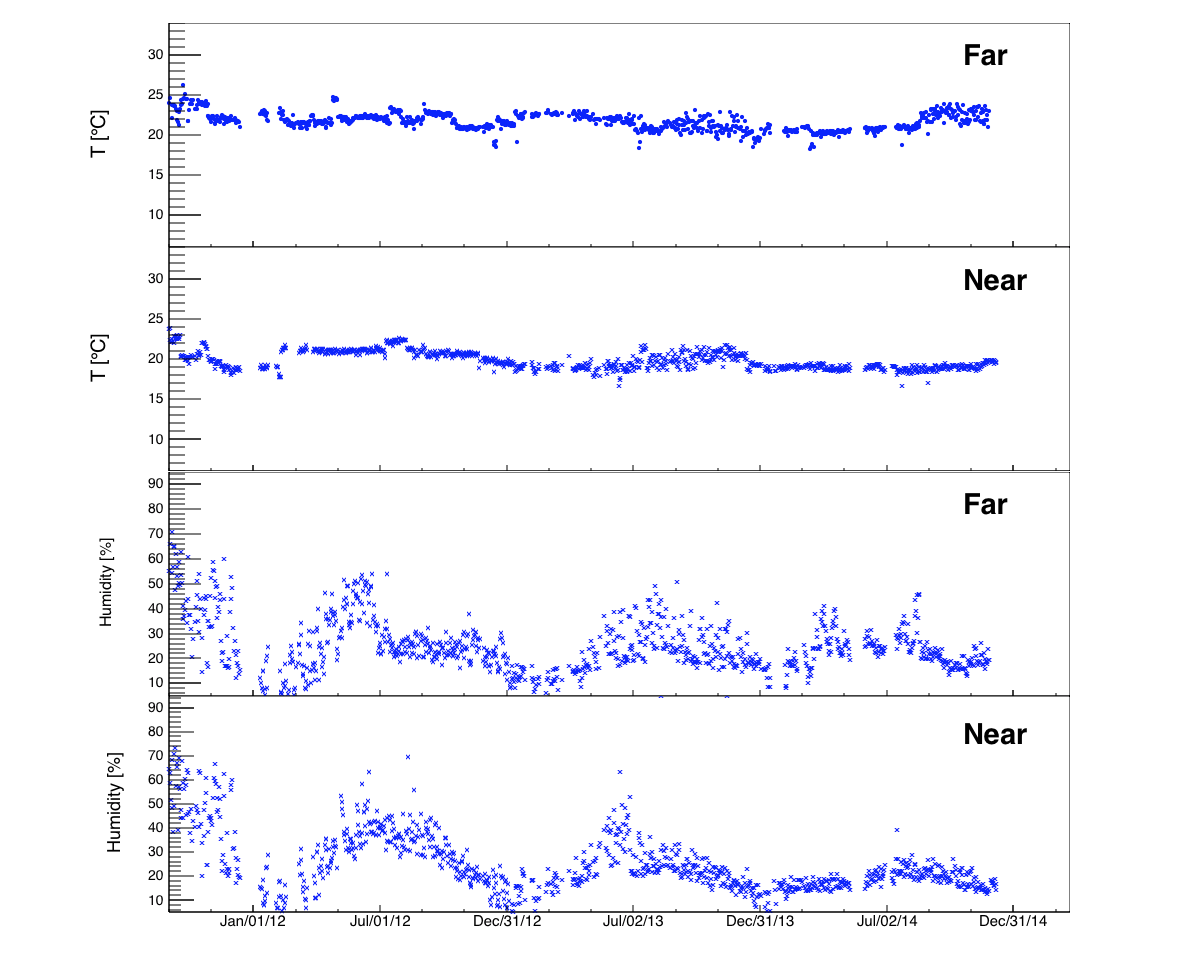}
\caption{Variation of temperature and humidity monitored at the far and near electronics huts. Seasonal variations of humidity are observed. Large fluctuations of humidity at small time intervals are caused by air currents from the air conditioner system.}
\end{figure}

\subsubsection{Radon Concentration} 
The radon monitor is connected to a PC with a 9-pin RS232 port. And the data is not provided to the online monitoring system.
Radon ($^{222}{\rm Rn}$) is a colorless and odorless gas produced by the radioactive decay of radium, which is abundant in common rocks such as limestone.  For monitoring $^{222}{\rm Rn}$ concentration in the experimental hall, the Sun Nuclear 1027 continuous radon monitor, which is based on a diffused junction diode, is used for safety. The radon monitor is directly connected to the SCS server with a 9-pin RS232 port, but the radon monitor does not provide the concentration data to the online monitoring system. 

The permissible $^{222}{\rm Rn}$ concentration in the air is less than 4 pCi/L or $0.148 ~{\rm Bq/L}$. From 4 pCi/L of radon gas, 1 pCi/L of radon can be dissolved into water at 22$^{\circ}$C, corresponding to one radon atom per 5.5 ${\rm cm^3}$ of water\cite{surbeck}. The radon in the OD water can produce background radiation for the reactor neutrino detection. The water circulation system can remove radioactive atoms from the water. The radon concentration in the far site experimental hall from March 2011 to August 2014 is shown in Fig. 9. Occasional high concentration readings can be understood by measurements during detector calibration using radioactive sources or ventilation off periods.

\begin{figure}[htbp]
\centering
\includegraphics[width=130mm]{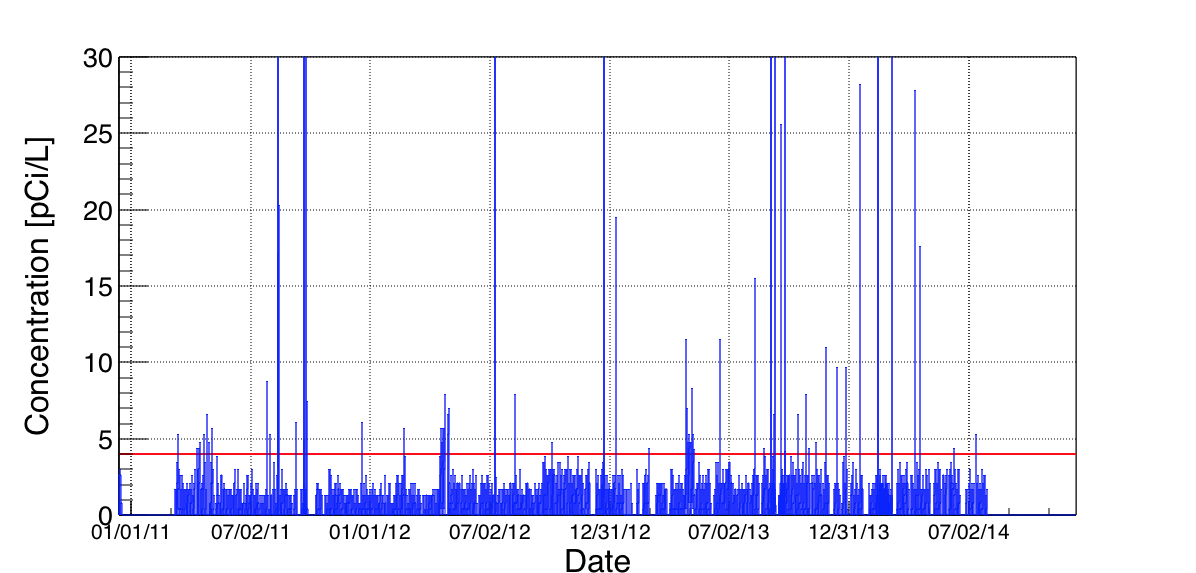}
\caption{Radon concentration inside the experimental hall at the far detector site. The occasional high dose is caused by the radioactive sources that are used for the calibration run. Solid line drawn across the plot shows 4 pCi/L. }
\end{figure}

\section{Summary}
The SCS are developed for acquiring and controlling the values of PMT HV and the water level in OD including the environmental parameters such as those of temperature, moisture and air conditions for safety management of the underground facilities. The SCS are based on commercial devices whose software libraries are designed for the {\sc Labview} platform and allow for continuous monitoring and stable operations of the RENO detectors. With the help of the SCS, RENO has acquired the inverse beta decay data using the two identical detectors successfully.

\section{Acknowledgments}
The RENO experiment is supported by the National Research Foundation of Korea (NRF) grant No. 2009-0083526 funded by the Korea Ministry of Science, ICT \& Future Planning. Some of us have been supported by a fund from the BK21 of NRF. We gratefully acknowledge the cooperation of the Hanbit Nuclear Power Site and the Korea Hydro \& Nuclear Power Co., Ltd. (KHNP).
We thank KISTI for providing computing and network resources through GSDC, and all the technical and administrative people who greatly helped in making this experiment possible. This work was supported, in part, by the NRF grant No. 2013R1A1A2011108.


\begin{thebibliography}{99}
\bibitem{jkahn1} J. K. Ahn {\it et al.,} (RENO Collaboration) Phys. Rev. Lett. {\bf 108}, 191802 (2012).
\bibitem{sbk} S. B. Kim,  Nucl. Part. Phys. Proc. {\bf 265-266}, 93 (2014).
\bibitem{jkahn2} J. K. Ahn  {\it et al.,} ``RENO: An Experiment for Neutrino Oscillation Parameter $\theta_{13}$ Using Reactor Neutrinos at Yonggwang'', arXiv:hep-ex/1003.1391 (2010).
\bibitem{isyeo} I. S. Yeo  {\it et al.,} Phys. Scripta {\bf 82}, 065706 (2010).
\bibitem{kspark} K. S. Park  {\it et al.,}  Nucl. Instrum. Meth. {\bf A686}, 91 (2012).
\bibitem{jspark} J. S. Park {\it et al.,} Nucl. Instrum. Meth. {\bf A707}, 45 (2013). 
\bibitem{caen} http://www.caen.it/.
\bibitem{LabVIEW} http://www.ni.com/labview/.
\bibitem{surbeck} Heinz Surbeck, Int. Conf. on Technologically Enhanced Natural Radioactivity caused by Non-Uranium Mining, Szczyrk, Poland (1996).
\end{thebibliography}
\end{document}